\documentclass[11pt]{article} % use larger type; default would be 10pt

\usepackage[utf8]{inputenc} % set input encoding (not needed with XeLaTeX)
\usepackage{amsmath}
\usepackage{geometry} % to change the page dimensions
\usepackage{graphicx} % support the \includegraphics command and options
\graphicspath{ {./pics/} }\usepackage{graphicx}
\usepackage{caption}
\usepackage{subcaption}

\usepackage{array} % for better arrays (eg matrices) in maths
\usepackage{verbatim} % adds environment for commenting out blocks of text & for better verbatim
\usepackage[procnames]{listings}
\usepackage{hyperref}

\newcommand{\be}{\begin{equation}}
\newcommand{\ee}{\end{equation}}

\title{\bf  Quintet Volume Projection}
\author{
    Vladimir Markov\\
    \texttt{vmarkov2@bloomberg.net}
    \and
     Olga Vilenskaia\\
    \texttt{ovilenskaia@bloomberg.net}
    \and
    Vlad Rashkovich \\
    \texttt{vrashkovich1@bloomberg.net} \\
\\
\large{ Bloomberg, L.P. in New York, NY}
}
\date{}
\begin{document}
\maketitle

%%%%%%%%%%%%%%%%%%%%%%%%%%%%%%%%%
%%%%%%%%%%%%%%%%%%%%%%%%%%%%%%%%%
% ABSTRACT
%%%%%%%%%%%%%%%%%%%%%%%%%%%%%%%%%
%%%%%%%%%%%%%%%%%%%%%%%%%%%%%%%%%

\begin{abstract}
We present a set of models relevant for predicting various aspects of intra-day trading volume for equities and showcase them as an ensemble that projects volume in unison.  
We introduce econometric methods for predicting total and remaining daily volume, intra-day volume profile (u-curve), close auction volume and special day seasonalities and emphasize a need for a unified approach where all sub-models work consistently with one another. 
Historical and current inputs are combined using Bayesian methods, which have the advantage of providing adaptive and parameterless estimations of volume for a broad range of equities while automatically taking into account uncertainty of the model  input components. The shortcomings of traditional statistical error metrics for calibrating volume prediction are also discussed and  we introduce Asymmetrical Logarithmic Error (ALE) to overweight an overestimation risk. 
\end{abstract}

%%%%%%%%%%%%%%%%%%%%%%%%%%%%%%%%%
%%%%%%%%%%%%%%%%%%%%%%%%%%%%%%%%%
% INTRODUCTION
%%%%%%%%%%%%%%%%%%%%%%%%%%%%%%%%%
%%%%%%%%%%%%%%%%%%%%%%%%%%%%%%%%%

\section{Introduction}
%\part{Introduction}

Accurate volume prediction is important for controlling and optimizing trading execution costs. 
All types of trading algorithms use volume prediction in order to choose the optimal trading rate  and volume execution trajectory (Markov, Mazur \& Saltz (2011)).
Traders must estimate volume to choose a proper execution algorithm and its front-end parameters, such as order duration or aggressiveness. Portfolio managers are aware that incorrect future volume estimation leads to excessive transaction costs  and poses serious constraints on capacity of an alpha-model. Efficient implementation of pre-trade TCA relies on volume prediction as well (Rashkovich \& Verma (2012)).

While most broker volume prediction algorithms are proprietary and not openly available to traders, there is a need for a generic model that can help traders  accurately select the front-end parameters of trading algorithms. 

There are a number of approaches to modeling intra-day volume.  
Brownlees, Cipollini \& Gallo (2011) proposed a dynamic model of intra-day volumes, taking into account one daily and  two intra-day components -- one periodic and one dynamic. 
Chen, Chen, Ardell \& Lin (2011) proposed a two-component hierarchical model combining  partial volume observed up to the time of prediction and dynamics of daily volume changes over time. 
Bialkowski, Mitchell \& Tompaidis (2014)  built a model of volume dynamics while developing a trading strategy that tracks VWAP using an autoregressive model for the logarithm of normalized volumes, including  explanatory variables such as stock-dependent time-of-day shape factor and a stock-independent day-of-week adjustment factor.
Calvori, Cipollini \& Gallo (2014) proposed a generalized autoregressive score (GAS) model for predicting volume shares,
taking into account an intra-day periodicity pattern and residual serial dependence. 
They assumed that the volume shares throughout the day follow a Dirichlet distribution with time-varying parameters. 
Satish, Saxena \& Palmer (2014) introduced a volume-forecast model dynamically weighting three components: rolling historical average for 15-minute bin trading volume;
per-symbol, per-bin ARMA (autoregressive moving average) model reflecting the serial correlation across daily volumes; and an additional ARMA model over deseasonalized intra-day bin volume data. 

Trading volume on option expiration dates compared to regular day for different markets was considered in Chiang (2009) for the U.S.; Gupta, Metia and Trivedi (2003), and Vipul (2005) for India;  Swidler, Schwartz and Kristiansen (1994) for Norway; and Corredor, Lechon \& Santamaria (2001) for Spain.

Despite their diversity, the above models have several common components. They combine the historical daily volume component with the intra-day component.  
They may have seasonal and dynamical sub-components. The autoregressive nature of the volume is captured by ARMA or ARIMA models.

We limit our model to continuous component of the trading volume.  The block-crossing component of volume requires a different approach (Glukhov (2007)).

%%%%%%%%%%%%%%%%%%%%%%%%%%%%%%%%%
%%%%%%%%%%%%%%%%%%%%%%%%%%%%%%%%%
% VOLUME PREDICTION METHODOLOGY
%%%%%%%%%%%%%%%%%%%%%%%%%%%%%%%%%
%%%%%%%%%%%%%%%%%%%%%%%%%%%%%%%%%

\section{Volume Prediction Methodology}

A practical model in finance is  always a compromise between mathematical rigor, underlying assumption about the data and practicalities of serving final user needs.  Instead of building one complex model, we construct five simple models (quintet) that work together in an ensemble to increase transparency and interpretability of the model.

The model combines: 
a historical total daily volume model (model one) that is based on the combination of the 20-day geometric average of the daily volume, ARMA component, and special days adjustments; an intra-day volume curve (u-curve) model (model two) that is based on the deep history curve (180 days), curve shift based on the overnight gap and expected total daily volume (functional regression); a close auction model (model three) that is based on  geometrical average and the seasonal adjustments for options expiration days. 

If the intra-day u-curve is stable, which is typical of liquid stocks, we use a Bayesian model that has a form of weighted sum of  historical daily volume component  and intra-day bin volume observations (model four). 
If the intra-day u-curve is unstable or noisy, which is typical of illiquid stocks, we use a model that has a form of weighted sum of  historical daily volume component and intra-day cumulative volume observation (model five).

As an error metric, we use asymmetric logarithmic error (ALE).
%
%The functional scheme is shown in Figure 1.
%
%\begin{figure}[!h]
%    \caption{Intra-day Volume Prediction Model}
%  \hspace{-1cm}
%  \includegraphics[scale=0.55]{Model_schema}
%\end{figure}

\section{Bayesian Inference} 

Financial data are inherently noisy, non-stationary and often have only finite sample leading to predictions characterized by large uncertainties. The Bayesian approach
gives a quantitative  framework for finding the best prediction despite that uncertainty, by assigning each possible state of the world a probability, and using the laws of probability to calculate the best prediction. 

Formally, Bayesian inference is an application of  Bayes' theorem to update the probability for a hypothesis as more evidence or information becomes available.  
In this framework, volume $V$ is a random variable that takes on a realized value $v$ once observed. 
$V$ is unobserved but described by some probability distribution that we want to derive from the actual data values $v$. Denote by $\theta$ parameters (such as the mean or the variance of a distribution) that characterize the probability model. The goal is to obtain estimates of the unknown parameters $\theta$ given the data $v$.

In Bayesian statistical inference, $\theta$ is random as well, possessing a probability distribution that reflects our uncertainty about the true value of $\theta$. 
Because both the observed data $v$ and the parameters $\theta$ are assumed to be random, we can model the joint probability of the parameters
and the data as a function of the conditional distribution of the data given the parameters and the
prior distribution of the parameters. 
\be
p(v,\theta)= p(v|\theta) p(\theta) , \quad p(v,\theta)= p(\theta|v) p(v) 
\ee
It leads to the famous  Bayes' theorem 
\be
 p(\theta| v) =\frac{p(v |\theta)p(\theta)}{p(v) }\propto p(v |\theta)p(\theta)
\ee
where $p(\theta|v)$ is referred to as the posterior distribution of the parameters $\theta$ given the observed data $v$, 
$p(v |\theta)$ is likelihood function, and $p(\theta)$ is the prior. The normalization factor doesn't depend on data parameters $\theta$  and  is given by:
\be
p(v)=\int\limits_{\theta} p(v|\theta) p(\theta) d \theta
\ee
The posterior probability is a function of a prior probability and a likelihood function that defines a statistical model for the observed data. Equation $(2)$  states that our uncertainty regarding the parameters of
our model, as expressed by the prior distribution $p(\theta)$, is weighted by the actual data via likelihood function $p(v | \theta)$ -- yielding an updated estimate of the model
parameters as expressed in the posterior distribution $p(\theta| v)$.

Although Bayes’  theorem is mathematically simple, its implementation can be computationally expensive. 
The difficulties lie in the normalizing constant $p(\theta)$, where the product of the prior and likelihood functions must be integrated over the valid domain of the parameters  being estimated. 
One way to get a tractable solution is to derive pairs of likelihood functions and prior distributions with convenient mathematical properties, including tractable analytic solutions to the
integral. Namely, if the posterior distributions $ p(\theta|x)$ are in the same family as the prior probability distribution $p(\theta)$, the prior and posterior are then called conjugate distributions, and the prior is called a conjugate prior for the likelihood function. 
Taking into account the log-normal approximation of the volume distribution, let us use well-known  results for conjugate priors and marginal distribution for normal random variables.

\subsection{Log-Normal Distribution  of Volume}

To formulate a model for Baysian inference, we need to specify the likelihood function.  Our research shows that log-normal distribution is a good fit for the main body of distribution for  both intra-day volume bins and daily volume. 
In Figure 1,  we show a typical example for a mid-cap stock.
\begin{figure}[!h]
%\centering
  \caption{Fitting Normal Distribution for log(volume) for ticker CAKE}
   \begin{subfigure}[b]{0.5\textwidth}
      \centering
      \includegraphics[scale=0.633]{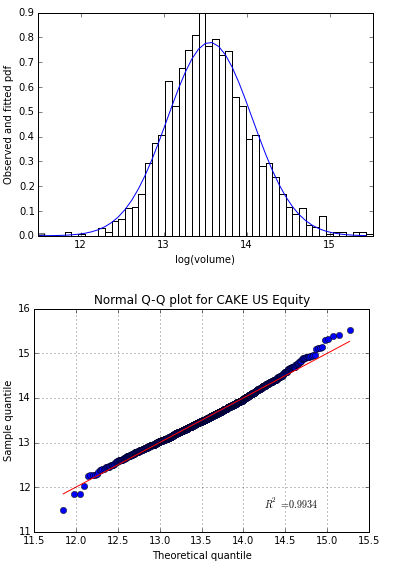}
      \caption{Daily data}
       %\label{fig:gull}
   \end{subfigure}
   \begin{subfigure}[b]{0.5\textwidth}
      \centering
      \includegraphics[scale=0.63]{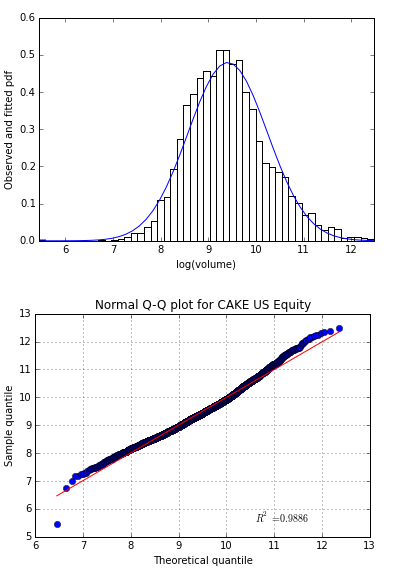}
      \caption{Intra-day data}
   \end{subfigure}
\end{figure}
Alternative distributions discussed in literature are q-gamma and Weibul distributions. 
Although tail behavior may vary, the analytic tractability and overall possession of many desirable properties (which we discuss later)  strongly favors a log-normal distribution as a base model for likelihood. The observations that  deviate from normality can be filtered out by the Grubbs filter (Grubbs [1950]). 

Assuming that volume follows a log-normal distribution means that the logarithm of volume $\log(V)\sim \mathcal{N}(\mu,\sigma)$ follows a normal distribution. 
The problem is to  estimate posterior distribution of the mean $\mu$ and standard deviation $\sigma$ of the expected daily volume given the finite number of intra-day bin volume observations and the prior information about past daily volumes.

Suppose that we are given data that is known to be independent and identically distributed
%from a normal process with known or unknown variance and unknown mean. 
and taken from a normal process with known or unknown variance and unknown mean. 
We wish to infer the mean of this process. There are two flavors of Bayesian inference in our case: the distribution has known variance $\sigma^2$ but unknown mean $\mu$; and the distribution has unknown variance $\sigma^2$ and unknown mean $\mu$.

\subsection{ Bayesian Inference with Unknown Mean and  Known Variance}

Suppose observations $D=\{x_i\}$ has known variance $\sigma^2$ but unknown mean $\mu$.
The posterior distribution of the mean $\mu$ is normal $P(\mu|D)=N(\mu|\mu_p,\sigma_p)$, with posterior mean of $\mu_p$ expressed as a weighted average of the sample mean $\bar x$ and the prior mean $\mu_0$ where the weights are proportional to precisions:

\begin{equation}
\mu_p=\frac{\frac{ n \bar x}{\sigma^2} +\frac{\mu_0}{\sigma_0^2}}{\frac{n}{\sigma^2} +\frac{1}{\sigma_0^2} } \,\, 
\end{equation}  
and posterior variance $\sigma_p^2$
\begin{equation}
\frac{1}{\sigma_p^2}=\frac{n}{\sigma^2} +\frac{1}{\sigma_0^2}  \,\, 
\end{equation}
here, $\bar x=\frac{1}{n}\sum_{i=1}^n x_i$ (see the Mathematical Appendix). 

Each observation increases the precision of the posterior distribution by the precision $\lambda=\frac{1}{\sigma^2}$ of one observation. The mean of the posterior $\mu_p$  is a convex
combination of the prior $\mu_0$ and the maximum likelihood estimator of the current daily volume $\bar x$ for Gaussian random variables $x_i$ , with weights proportional to the relative precisions.

If we are interested only in inferences about the mean, and if the sample size is not too small, we can get a reasonable approximation of the posterior distribution by treating the standard deviation as known and equal to the sample standard deviation. 
A more accurate representation of our knowledge should account for the unknown variance.

\subsection {Bayesian Inference with Unknown Mean and Unknown Variance}
The conjugate prior for mean $\mu$ and precision $\lambda=1/\sigma^2$ is normal gamma distribution. The marginal distribution of the mean $P(\mu|D)$ (given observations $D=\{x_i\}$) is given by the Student's t-distribution and the estimation of the mean is a simple average between prior $\mu_0$  and average of $n$ observations $\bar x=\frac{1}{n}\sum_{i=1}^n x_i$.
\begin{equation}
\mu_p=\frac{\mu_0 \kappa_0+n \bar x}{\kappa_0+n},
\end{equation} 
where $\kappa_0$  parameter is the effective size of the prior sample.  The natural values of $k_0$ are $k_0 \in [0.3-0.8] N_{prior}$, here $N_{prior}$ is the number of observations of the historical daily volume (in our case $N_{prior}=20$ and the bin size is ten minutes).

\section{Volume Prediction Error Metric}

The volume prediction error metrics have to take into account the asymmetric risk profile of execution and fat tails of volume distribution. 
For example, overestimation  of the daily volume by a factor of two leads to the same increase in participation rate. 
If there is an obligation to complete the order and the target participation were 20 percent, the actual participation would be 40 percent, leading to excessive market impact.
The conservative estimation of future volume gives more freedom to an execution algorithm and leads to impact savings, especially for orders with slow alpha decay. 
Also, the log-normal distribution of volume possesses fat tails and the error metric has to be robust to handle large-volume days. 

To calibrate model parameters, we use Weighted Asymmetrical Logarithmic Error (ALE): 

\be
\text{ALE} = \sum_{i=1}^n w_i  (X_i^{\text{est}} - X_i^{\text{true}}) \cdot \left | X_i^{\text{est}} - X_i^{\text{true}}\right |,
\ee
where 

\be
w_i(x)=\begin{cases}
	1, \text{  if  } x \le 0\\
	2, \text{  if  } x > 0
	\end{cases}
\ee
and $X_i = \log(V_i)$.
Thus, ALE is, in fact, an asymmetrical generalization of $L_1$ norm in logarithmic space.
In ALE, we use double weights for overestimation errors in order to take into account the asymmetric profile of  risk of execution.

We note that Root Mean Square Error (RMSE) can be influenced by tail days and is symmetric, and that  Mean Absolute Percentage Error (MAPE) metric is risk symmetric as well.  
The $R^2$ metric has limited value outside of the linear regression framework. 

To illustrate the practicality of our approach, we use representative samples from the S\&P 500, S\&P Midcap 400 and Russell 2000 indexes. 
For our test sample, we sort names by index weight and take half of the top 100 names from S\&P 500, 100 mid-range names from S\&P Midcap 400 and 100 bottom names from Russell 2000 names. 
The data range is  between July, 2015 and December, 2016.

\section {Quintet Volume Projection}

\subsection {Volume Prior}

Without any intra-day information, the historical daily volume and its averages are often used as a proxy for today's expected volume. 

The 20-day moving average is an industry standard for estimating liquidity of a stock.  
Moving averages have a tendency to overestimate volume due to memory of large-volume days that happen quite frequently. 
Corporate news, earnings announcements, option expirations and index rebalancing all can lead to large-volume days. 
Formally, large-volume days lead to the fat right tail of daily volume distributions.  

There are multiple views on average. 
Subjectively, it is the number in the middle or a number that is balanced. 
Another goal when applying the average is to understand a data set by using a single  representative number. 
Mathematically, for many distributions of numbers there is a native average; the calculation of average depends on the distribution. 
The arithmetic mean of normal random variables is normal, the geometric mean of log-normal random variables is log-normal, while Cauchy random variables are closed under taking harmonic means. 

Since it's more risky to overestimate volume than underestimate it, the geometric average gives a better performance in ALE metrics than does the arithmetic mean. 
Also, the geometric mean of a log-normal random variables is equal to its median, which represents a typical trading volume and is not influenced by outliers.

As a simple prior estimator of logarithm of total daily volume, an average of the most recent $N=20$  log-daily volume observations $X_i=\log(V_i)$ 
is taken $\mu=\frac{1}{N}\sum_{i=1}^{N+1} X_{t-i}$. In physical space, the prior is $V_{prior}=e^\mu$ and is given by a  geometric mean (GM) of volume: 
\begin{equation}
GM[V]=e^\mu=e^{\frac{1}{N}{\sum_{i=1}^N X_i}} =(V_1\cdot V_2\cdot ...\cdot V_N)^{1/N}
\end{equation}
We note that given the log-normal distribution with parameters $\mu$ and $\sigma$,  the arithmetic mean is given by: 
\begin{equation}
E[V]=\frac{1}{N}(V_1+V_2+...+V_N)=GM[V] \cdot e^{\sigma^2/2}
\end{equation}
  
According to a well-known inequality concerning arithmetic and geometric means for any set of positive numbers,
the arithmetical average is always greater or equal to the geometrical average.

\paragraph*{Ajusting Prior for ARMA Component}

Autoregressive–moving-average (ARMA) models provide a description of a stationary stochastic process in terms of two polynomials, one for the autoregression and the second for the moving average. 
The notation ARMA(1,1) refers to the model with one autoregressive term and one moving-average term:
\be
y_{t}=\varphi  \, y_{t-1}+\varepsilon _{t}+\theta\, \varepsilon _{t-1}
\ee
Here, $y_t = X_t - \mu_t$, 
$X_t = \log(V_t)$ -- logarithm of total daily trading volume of day $t$, 
$\mu_t$ is $N=20$-day moving average:
$\mu_t=\frac{1}{N}\sum_{i=1}^{N} X_{t-i} $. In this section, subindex $t$ refers to the index of an historical day.To make a volume series (quasi) stationary we substract from log-volume observations $X_t$ the running average $\mu_t$:  $y_t = X_t - \mu_t$.
The coefficients $\varphi $ and $\theta$ were estimated per stock by minimizing ALE metrics between realized and estimated volume. 
The fitted parameters are almost universal for the S\&P 500 names and only slightly vary by stock, with typical value given by $\varphi \approx 0.7$ and  $\theta \approx -0.3$.  More complex models of ARIMA$(n,p,q)$ class do not give significant risk metric improvement over ARMA$(1,1)$.

%The lagged values of $y_t$ that appear in the equation are called “autoregressive” (AR) terms, and the lagged values of the forecast errors are called “moving-average” (MA) terms. 
The model contains the AR(1) and MA(1) models; it is instructive to look at  MA(1)  and AR(1) processes separately to better understand the dynamics that they describe. 
For an AR(1) process $x_t =  \varphi\, x_{t-1} + \varepsilon _{t}$ with $ | \varphi | < 1$, the effects of $  \varphi$ on $x$ are:
\be
\begin{tabular}{l l l l l l}
$\varepsilon$:& 0& 1& 0& 0& ...\\
$x$ :& 0& 1 &$\varphi$ & $\varphi^2 $&...\\
\end{tabular}
\ee

The AR(1)  model describes autoregressive  behavior in which next period’s value should be predicted to be $\varphi$ times as far away from the mean as the previous period’s value.

For an MA(1) process, $x_t = \varepsilon _{t} + \theta \,  \varepsilon_{t-1}$, the effects of $\theta$ on $x$ are:
\be
\begin{tabular}{l l l l l l}
$\varepsilon$ :& 0& 1& 0& 0& 0\\
$x$ :& 0& 1 &$\theta$& 0 &0\\
\end{tabular}
\ee
The lagged values of the forecast errors are called moving-average (MA) terms.  After a big volume day where error of prediction $y_t>0$ is positive,  the MA component dumps the prior for the next day due to a negative sign of $\theta$.

In essence, the MA part models the response of the market to external shocks such as earning announcements or significant corporate news; endogenous trends of volume are modeled by the AR component.

An example of ARMA prediction versus 20-day arithmetical average and 20-day geometrical average predictions is shown in Figure 2.

\begin{figure}[!h]
    \caption{Arithmetical average, geometrical average and ARMA predictions for IBM US Equity}
  \hspace{2cm}
  \includegraphics[scale=0.9]{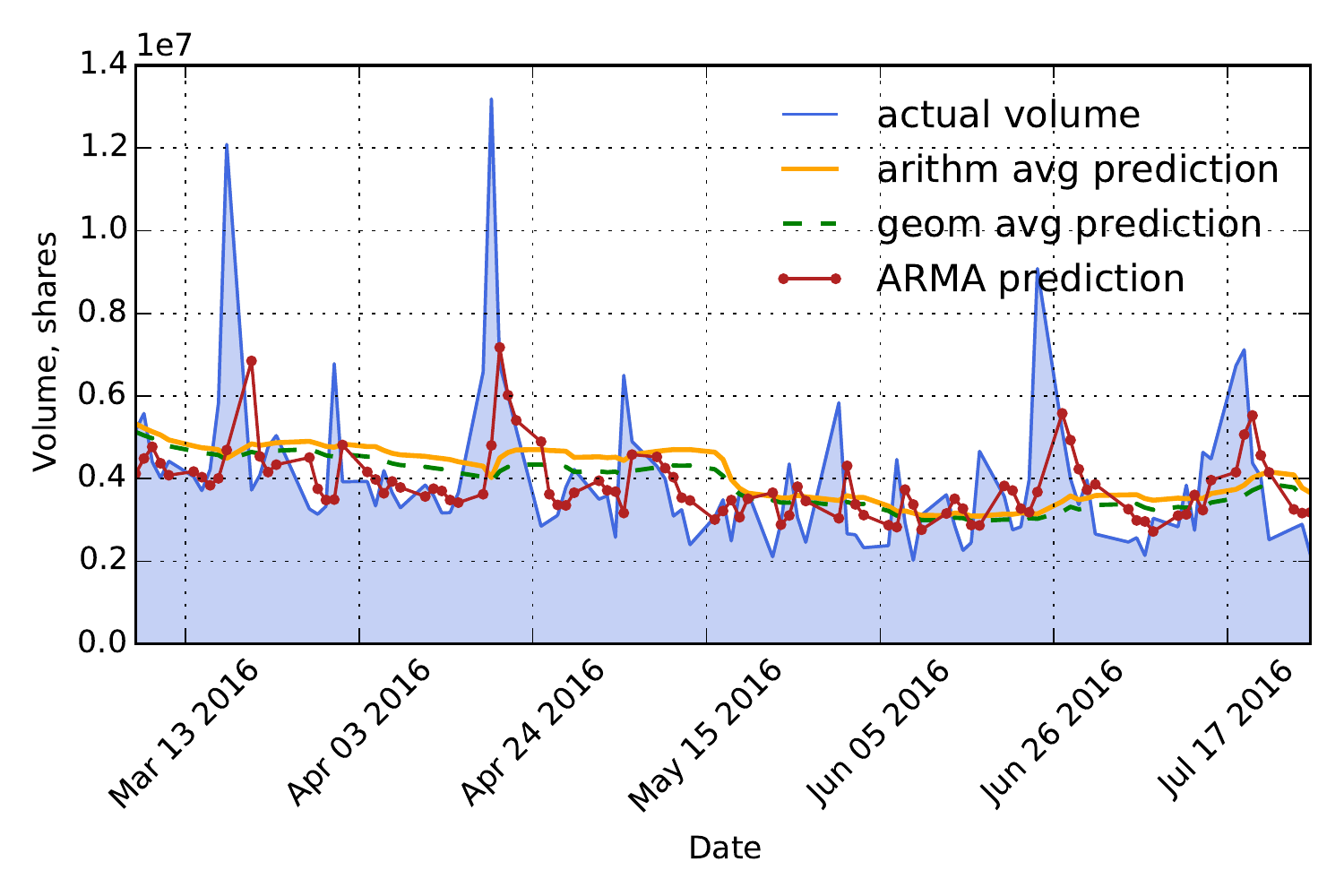}
\end{figure}

\paragraph*{Adjusting Prior for Special Days}

Special days -- such as earnings announcements, options expirations, index rebalancing, high overnight price gaps, etc.-- may be informative and often lead to higher trading volume.
To take this into account, a linear regression with ALE error metrics can be performed:
\be
y_t =\sum\limits_{k=1}^{m} \beta_k x_{k,t} + \varepsilon_t, \qquad \varepsilon_t \sim \mathcal{N}(0, \sigma^2_{\varepsilon})
\ee
Here, $y_t = X_t - \mu_t$, $X_t = \log(V_t)$, $\mu_t$ is $N=20$-day moving-average $\mu_t=\frac{1}{N}\sum_{i=1}^{N+1} X_{t-i}$, and m is the number of independent variables.

The choice of  dependent variable  $y_t = X_t - \mu_t$ allows easy interpretation of the regression coefficient $\beta$ as a multiplier for the prior. 
If the only independent variable $x_{1,t}$ in the regression is the overnight price gap $ g_t$, then the gap multiplier $\eta_{gap}=\exp(\beta g_t)$ for today's volume is given by:
\be
V_t=\exp(\mu_t) \times \eta_{gap}
\ee

 Figure 3 shows a cross-sectional regression (14) of excess logarithmic volume on overnight price gap to volatility for representative sample of  S\&P 500 index.
%\footnote{50 securities are randomly chosen from the index}

\begin{figure}[!h]
    \caption{Cross-sectional regression (14) of excess logarithmic volume on overnight price gap to volatility ratio for S\&P 500 index sample}
  \hspace{2cm}
  \includegraphics[scale=0.9]{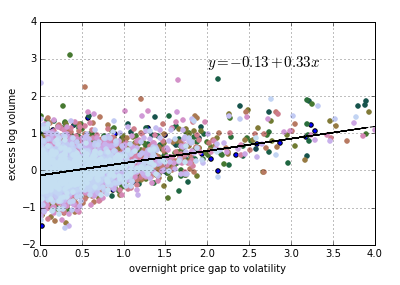}
\end{figure}

\subsection{Intra-day Volume Profile: U-curve}

The separate modeling  of the total daily volume level from its intra-day shape (u-curve)  increases stability and interpretability of the model.  Moreover, the u-curve has its own value as it is used by VWAP-type algorithms. 

We define intra-day profile $u(t)$ (u-curve) as the fraction of the day's volume that has been traded during $i$-th bin at time $t$.
The cumulative sum of u-curve $c(t)$(c-curve) is the fraction of the day's volume that has been traded from market open until time $t$:
\be
c(t)  = \frac{V(t)}{V(T)} , \qquad u(t)=c(t)-c(t-1)
\ee
Here,  $V(T)$ is the total daily trading volume and $V(t)$ is the total volume traded up to time $t$. In this section, subindex $t$ refers to the index of an intraday-bin.
The volume curves $c(t)$ and $u(t)$ are known only after the close, so the estimated $\hat c(t)$ and $\hat u(t)$ have to be used to make predictions.

The intra-day volume profile tends to be stable and on average does not change significantly over time. A plain approach to obtain an u-curve estimator is to take an average u-curve estimated over a prolonged historic period, e.g. the last 180 days.

\paragraph*{Functional Regression for U-curve.}

Functional Data Analysis (FDA) deals with the analysis and theory of data that are in the form of functions. In functional regression,  responses or covariates are functional or vector data.  In this section, we model dependence of cumulative u-curve $c(t)$ on overnight price gap and  quantile level of daily volume: 

%  on It allows to model the shape of u-curve dependence on overnight price gap or quantile level of daily volume. The regression is 
\begin{equation}
c_i(t) =\beta_0(t)+ \sum\limits_{k=1}^{m} \beta_k(t) x_{k,i} + \varepsilon_i(t), \qquad \varepsilon_i(t) \sim \mathcal{N}(0, \sigma^2_{\varepsilon})
\end{equation}
here $i = 1, ..., n$ - $i$-th day from $n$ historical days available; $t = t_0, ..., T$ - $t$-th bin of current day; m is the number of independent variables;
$x_{k,i}$ -- $k$-th independent scalar predictor on day $i$;
$\beta_k(t)$ - partial effect of predictor $x_k$  on the response at time $t$.

For visualization purposes, we perform two separate functional regressions: for overnight price gap and for the total daily trading volume percentile.

In the first regression, we use the overnight price gap as an independent external parameter $x_k$. The gap is defined as the relative difference between the current day open price and the previous day closing price over the 20-day price volatility.
According to the regression  (17) for the vast majority of securities, the intra-day volume profile has higher values in the beginning of the day for days with higher ratios of overnight price gap to price volatility.
It means that for high-overnight-price-gap days, the intra-day volume profile changes from U-shape closer to inverted J-shape.
Figure 4 shows average cumulative intra-day volume profiles for different values of overnight price gap to price volatility for the S\&P 500, S\&P Mid-cap 400 and Russell 2000 indexes representative samples.
%\footnote{50 securities are randomly taken from each index to create the charts.}

\begin{figure}[!h]
  \caption{Cumulative volume profiles by overnight price gap to volatility for the S\&P 500, S\&P Midcap 400, and Russell 2000 indexes samples}
   \begin{subfigure}[b]{0.3\textwidth}
      \centering
      \includegraphics[scale=0.32]{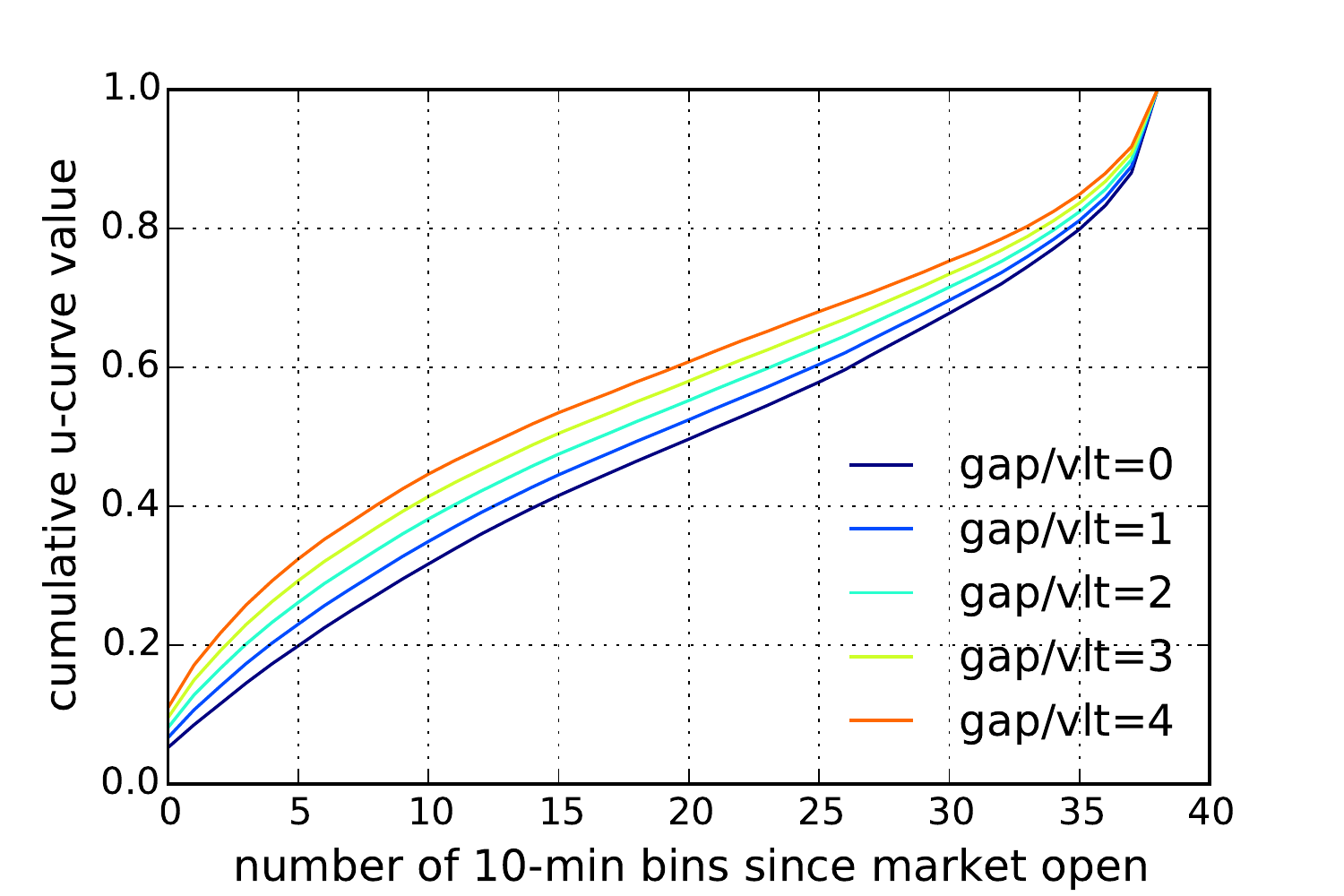}
%      \includepdf[scale=0.45]{c_curves_gap_spx_notitle.pdf}
      \caption{S\&P 500}
   \end{subfigure}
   \begin{subfigure}[b]{0.3\textwidth}
      \centering
      \includegraphics[scale=0.32]{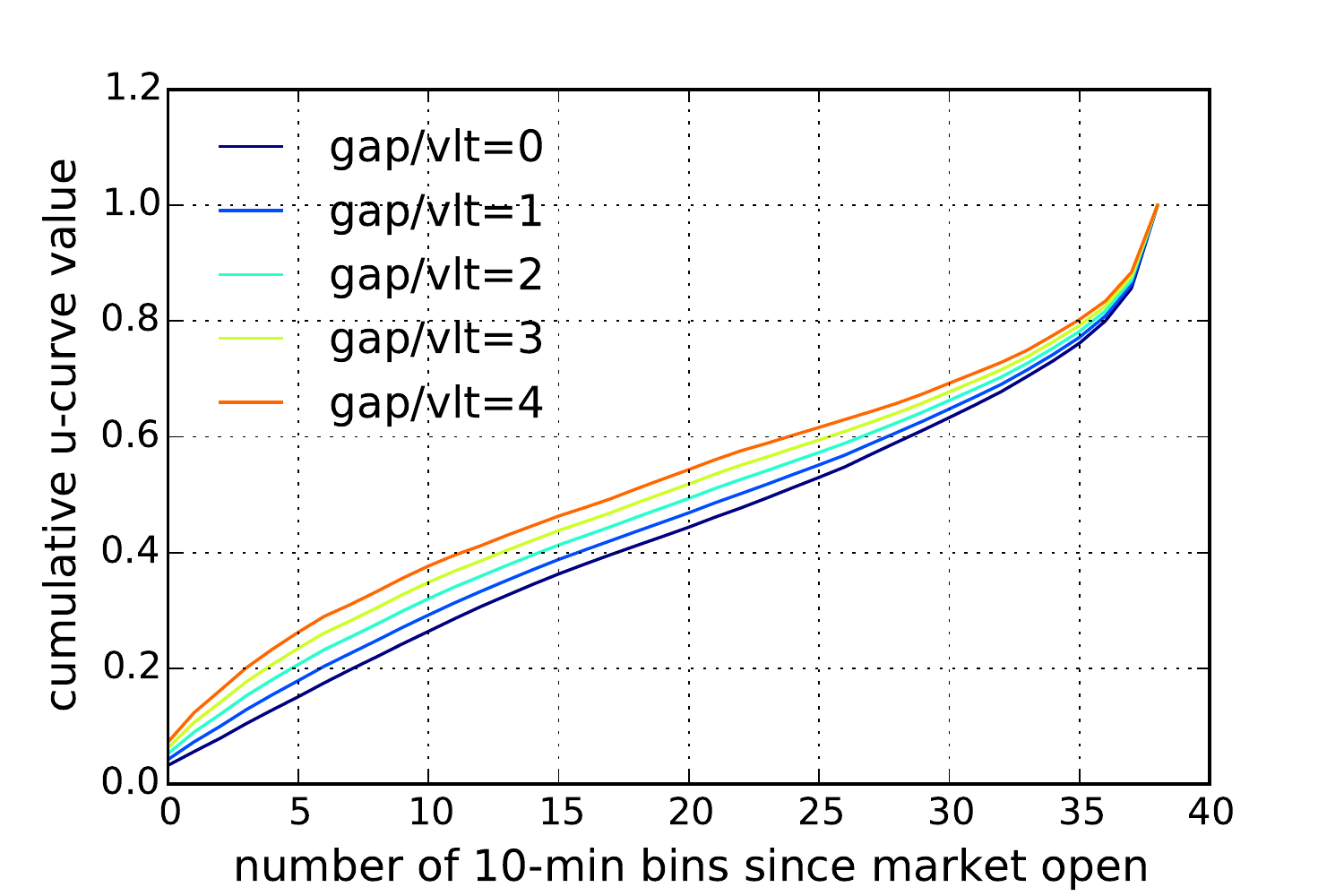}
      \caption{S\&P Midcap 400}
   \end{subfigure}
   \begin{subfigure}[b]{0.3\textwidth}
      \centering
      \includegraphics[scale=0.32]{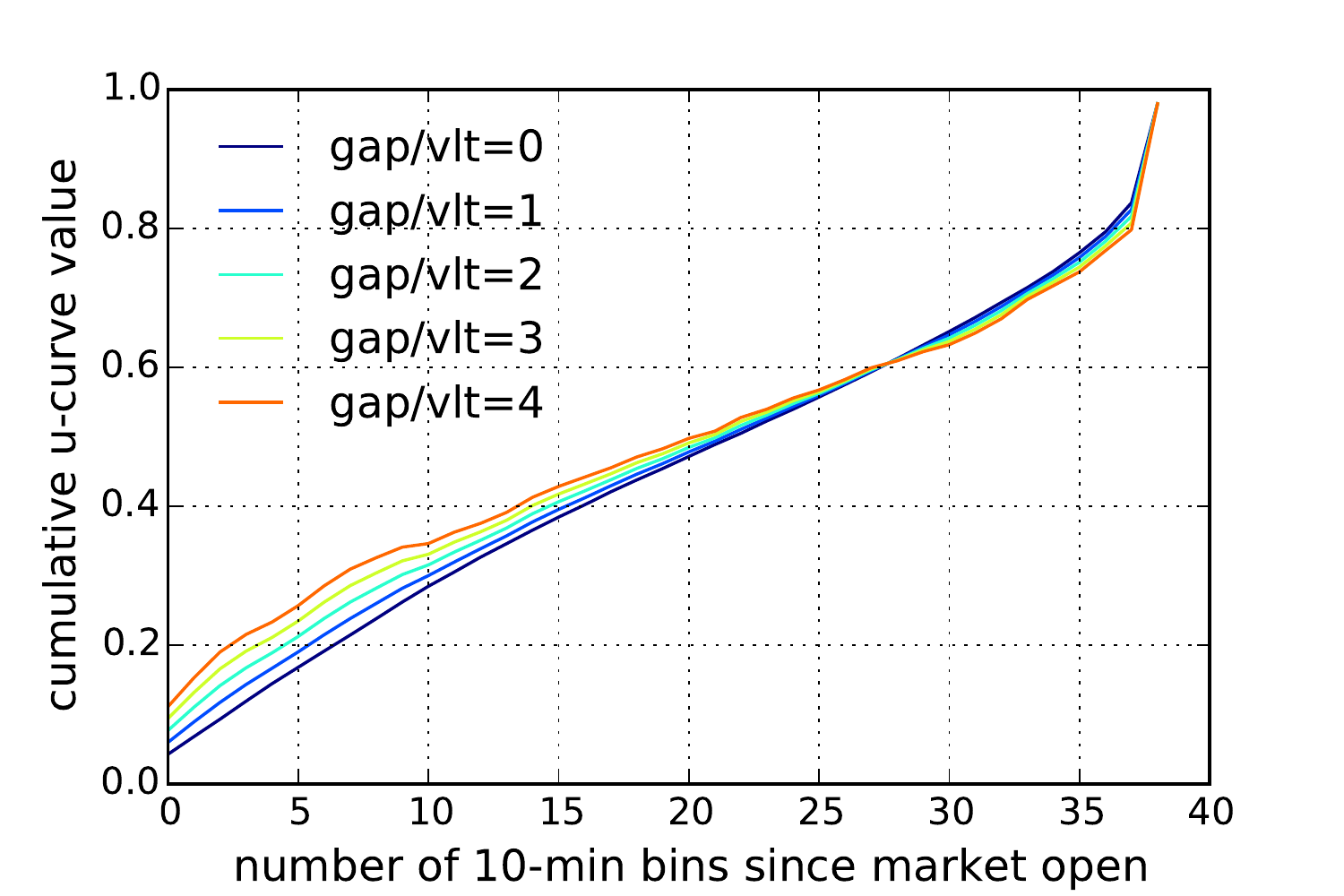}
      \caption{Russell 2000}
   \end{subfigure}
\end{figure}

In Figure 5, there are average coefficients of the functional regression (17) for S\&P 500 index representative sample
%\footnote{50 securities are randomly chosen from the index}
on a single independent variable - overnight price gap:
\begin{equation}
c_i(t) =\beta_0(t)+ \beta_1(t) g_i + \varepsilon_i(t), \qquad \varepsilon_i(t) \sim \mathcal{N}(0, \sigma^2_{\varepsilon}),
\end{equation}
here $g_i$  is the overnight price gap to volatility on day $i$. 
Higher value of $\beta_1$ in the beginning of the trading day means that a  higher overnight price gap results in a higher volume rise for the first bins of the trading day than for the last ones.

\begin{figure}[!h]
    \caption{Average coefficients of functional regression for S\&P 500 index}
  \hspace{4cm}
  \includegraphics[scale=0.6]{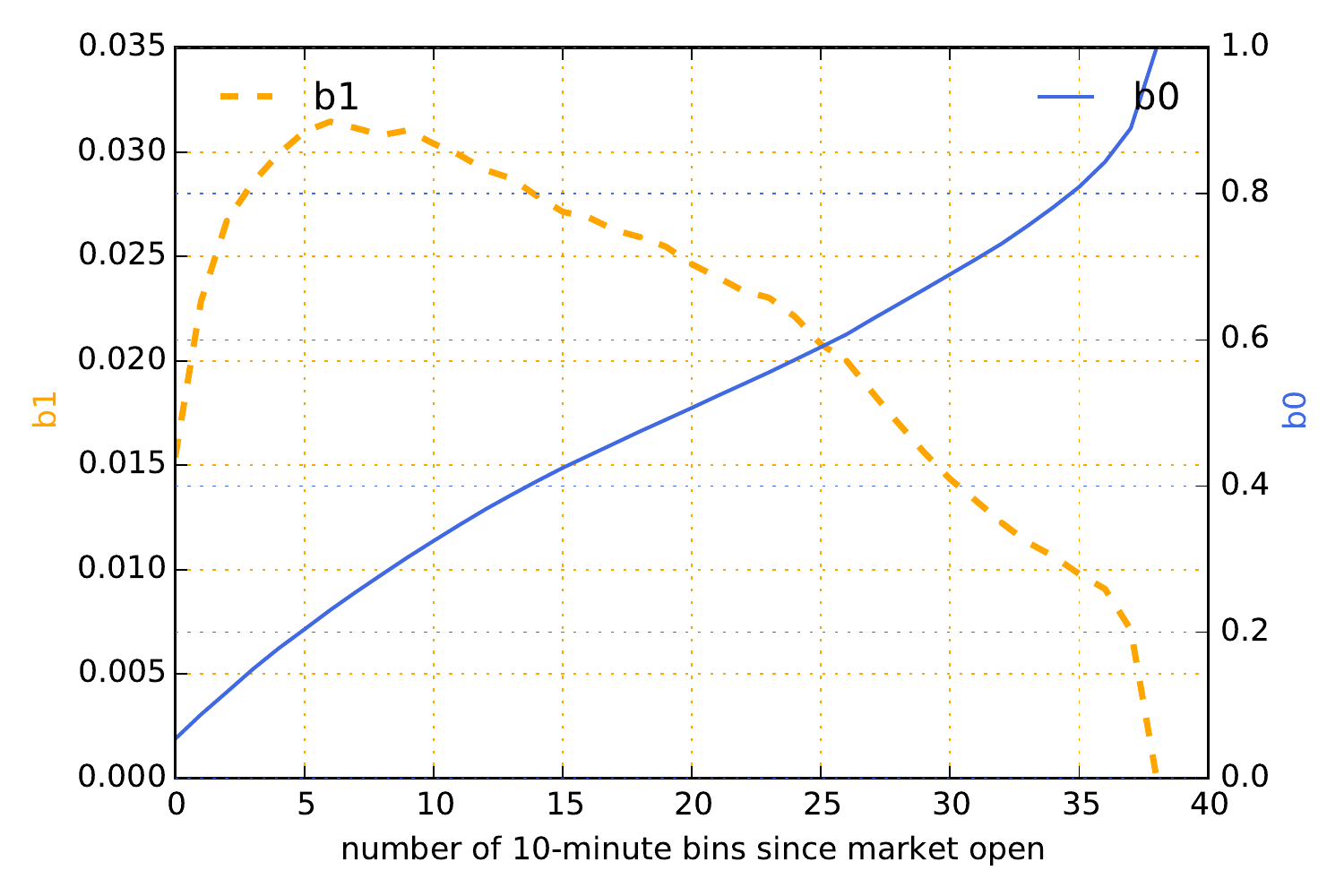}
\end{figure}

Similar to the regression on overnight price gap, the regression on total daily trading volume percentile
(the percentage of historical days for which total daily trading volume is lower than for the current day)
indicates that for high-volume days the intra-day volume profile changes from U-shape closer to an inverted J-shape.
Figure 6 offers average cumulative intra-day volume profiles for different values of total daily trading volume percentiles for S\&P 500, S\&P Midcap 400 and Russell 2000 indexes representative samples.

\begin{figure}[!h]
  \caption{Cumulative volume profiles by total daily trading volume percentile for the S\&P 500, S\&P Mid-cap 400, and Russell 2000 indexes}
   \begin{subfigure}[b]{0.3\textwidth}
      \centering
      \includegraphics[scale=0.32]{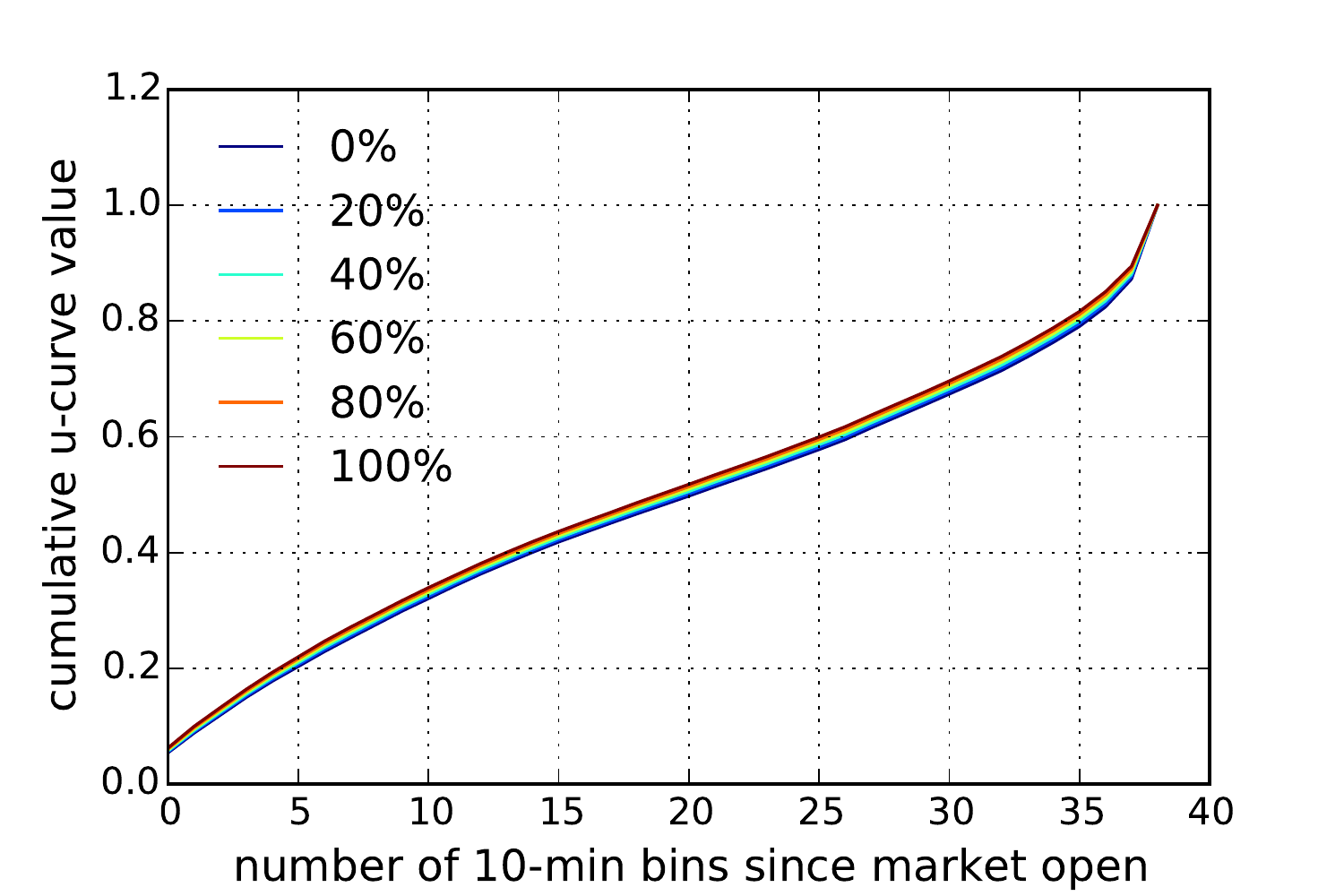}
      \caption{S\&P 500}
   \end{subfigure}
   \begin{subfigure}[b]{0.3\textwidth}
      \centering
      \includegraphics[scale=0.32]{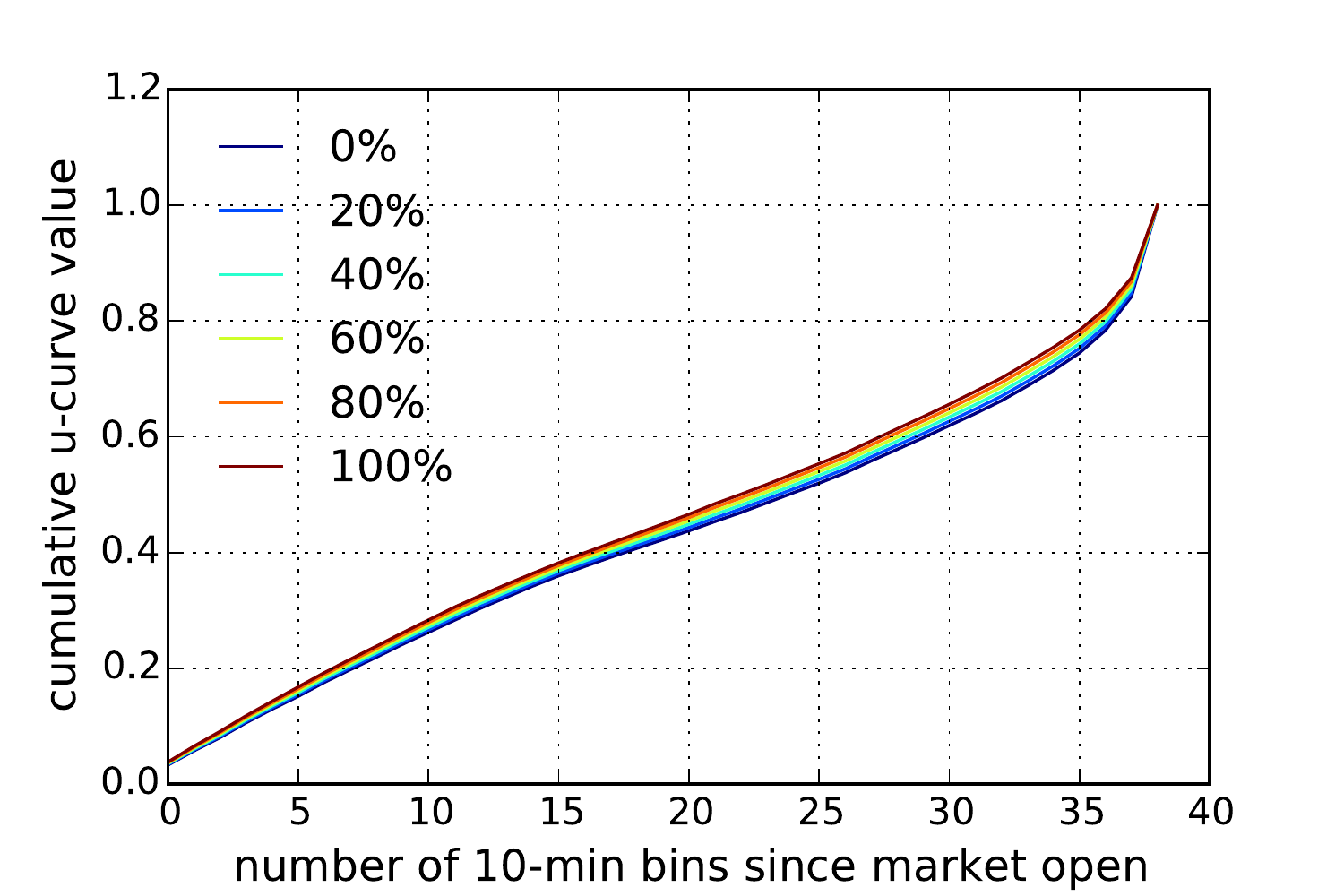}
      \caption{S\&P Midcap 400}
   \end{subfigure}
   \begin{subfigure}[b]{0.3\textwidth}
      \centering
      \includegraphics[scale=0.32]{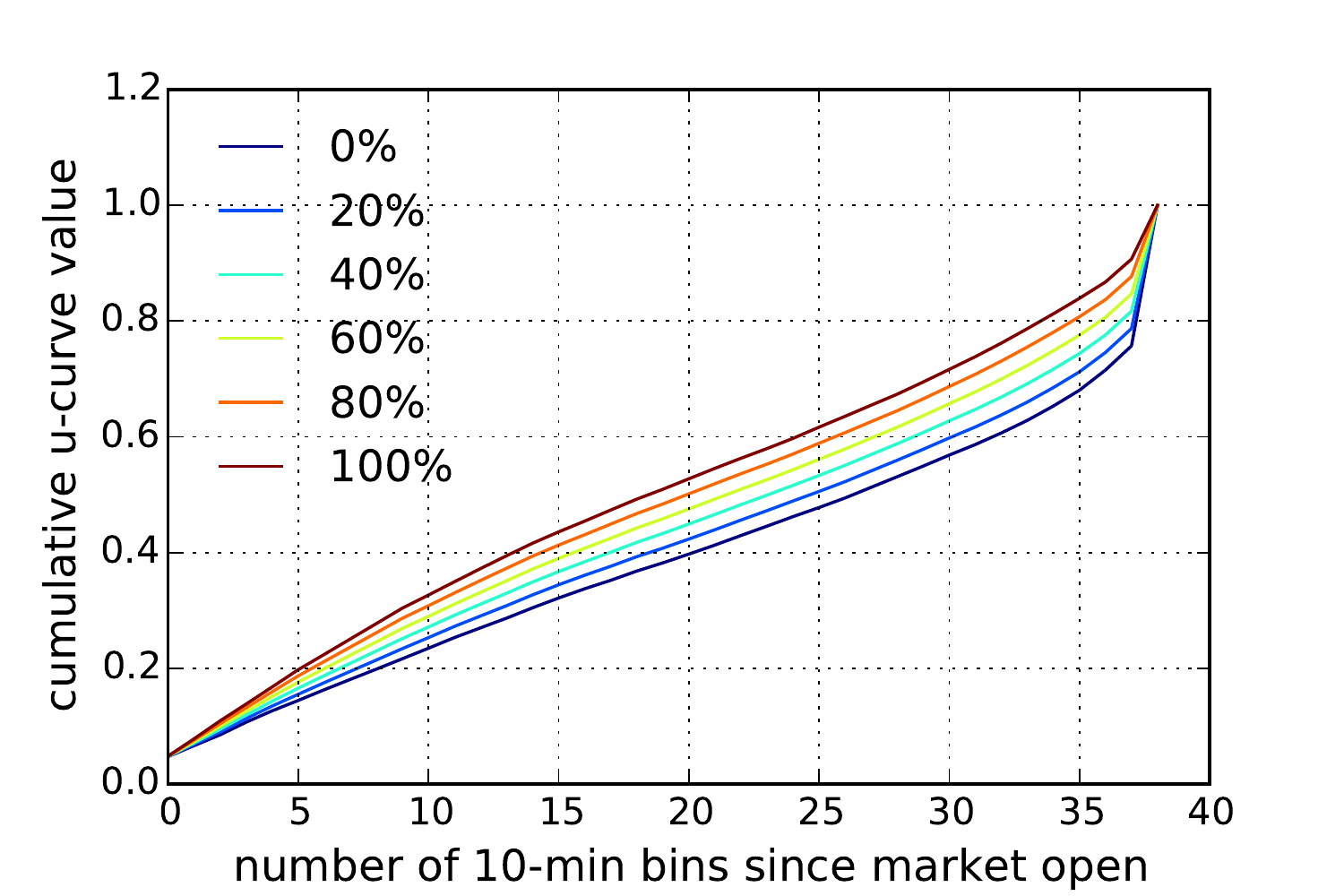}
      \caption{Russell 2000}
   \end{subfigure}
\end{figure}

The dependence of  the shape of the u-curve on total daily volume can be used to update the shape of u-curve intra-day based on total daily volume predictions.

Although we plot the aggregated results only, the regression (17) coefficients  are often stock-dependent.

\subsection{Volume Model for Liquid Securities: Intra-day Bin Model}

For liquid securities, the intra-day prediction  for the daily volume  $x(j)$ based on observation of volume traded in bin $j$  $v(j)$ is given by:
\begin{equation}
x(j)=\log \left(\frac{v(j)}{\hat u(j)}\right)
\end{equation}  
Note that we use the estimated $u$-curve $\hat u$, not the true one which becomes known only after the day closes. It shows the value of modeling $u$-curve discussed in the previous section.

In the beginning of the day, the variance of intra-day observations is unknown and the log-daily volume after n bins is estimated using formula (6):

\begin{equation}
\mu(n)=\frac{\mu_0 \kappa_0+n \bar x}{\kappa_0+n},
\end{equation} 
where $\kappa_0$  parameter is the effective size of the prior sample.  The optimal value of $k_0$ is bin size-, market cap- and country-dependent. The natural values of $k_0$ are within range  $k_0 \in[0.3-0.8] N_{prior}$. For example, for liquid names in the U.S. and 10-minutes bins, we use $k_0=0.5$.   After there are enough observations to estimate the variance  $\Sigma^2$  of intra-day observations $x$, 
formula (4) is used and  the  estimated log-daily volume is given by:
\begin{equation}
\mu(n)=\frac{\frac{n \bar x}{\Sigma^2} +\frac{\mu_0}{\sigma_0^2}}{\frac{ n}{\Sigma^2} +\frac{1}{\sigma_0^2} }
\end{equation}  

Here $\mu_0$ and $\sigma_0^2$ are mean and variance of the logarithmic prior,
and  $ \bar x$ and $\Sigma^2$ are mean and variance of n current-day estimates $x(n)$ defined by the Formula (19).

\subsection{Volume Model for Illiquid Securities: Historical Cumulative Model}

There may be cases when calculating total daily volume estimators based on each particular bin is not optimal (e.g., when the data is sparse, not very stable, etc.)

Illiquid stocks often have bars with zero volume, making u-curve erratic. In such case, use of the cumulative curve for forming intra-day observations looks more promising. 
Let's define as $z(i)$ the log of daily volume based on the estimated cumulative u-curve $\hat c(i)$ and cumulative intraday volume up to time $i$ $V(i)$ 
\begin{equation}
%z_j=\log \left(\frac{V_j}{ c_{j}}\right)\approx \log \left(\frac{V_j}{\hat c_{j}}\right)
z(i)=\log \left(\frac{V(i)}{\hat c(i)}\right)
\end{equation}  

The estimate of the log-daily volume in the historical cumulative model after $n$ bins is given by: 
\begin{equation}
\mu(n) =\frac{\frac{\mu_0}{\sigma_0^2}+\frac{z(n)}{\bar \Omega^2(n)}} {\frac{1}{\sigma_0^2}+ \frac{1}{\bar \Omega^2(n)}} 
\end{equation}  
Here $\Omega^2(n)$ is a dispersion of daily prediction on time $n$ using $z(n)$ over the last $M$ days: 
\begin{equation}
\Omega^2(n)=\frac{1}{M} \sum_{I=1}^M (z^{(I)}(n)-X^{(I)}-\overline{(z^{(I)}(n)-X^{(I)})})^2 
\end{equation}  
here $X^{(I)}$ is the total daily volume of day $I$.
\subsection{Auction Volume Prediction}

The volume transacted at the closing auction represents an important and significant fraction of the daily volume. Both absolute auction volume  and the volume at the close measured as a proportion of the total volume of the day are highly volatile and hard to predict.

The close auction price is defined by the price that maximizes the number of crossed shares. Given that the order size submitted by a trader is small relative to the typical auction volume (that  minimizes the risk to affecting the closing price) and the rest of the order flow is random, there is a high probability that the auction price will be near to the close price of the continuous session. A number of traders use close price as a benchmark and any deviation from it represents a risk for them. 

The close auction allocation has to be decided  in advance and thus the simplest (and most robust) strategy is to submit a fixed percentage of the predicted volume.  
%http://fixglobal.com/home/the-12-rule-for-asias-closing-auctions/
Stone, G., T. Kingsley, G. Kan [2015] recommend following a particular "rule of thumb”  to minimize the price impact during the close auction: take the lesser of 12 percent of the predicted closing auction volume and 12 percent of your order and allocate that share amount to the closing auction.
ALE metrics provides a trade-off between a reasonable prediction quality and risk of overestimation that matches the objectives of the fixed percentage strategy.

As a base prediction of the close auction volume, we take 20-day geometric average. In some cases the ALE error can be slightly (within 5  percent) improved by an ARMA model. Unfortunately, the ARMA signal is weak and doesn't justify the increase in complexity of the model. 

In general, the close-auction volume increases around major option and future expirations and rolls and index rebalancing.

In the U.S., the most noticeable spike of auction volume is seen during triple witching days. Triple witching day is the third Friday of every March, June, September, and December. On those  days, the market experiences the simultaneous expirations of stock market index futures,stock market index options and stock options.

%Auction volume shows strong seasonality effects around triple witching days. 
%For the U.S. these are third Fridays of March, June, September and December. 
Figure 7 gives  a typical example of closing auction volumes. 

\begin{figure}[!h]
    \caption{Closing Auction Volume for IBM US Equity}
  \hspace{2cm}
  \includegraphics[scale=0.9]{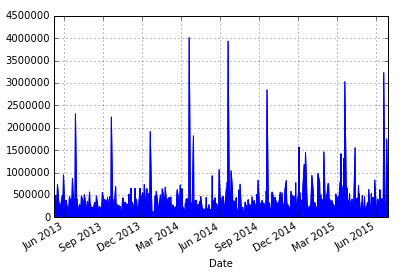}
\end{figure}

To take the seasonality into account, we perform a linear regression: 
\be
y_t =  \beta d_t + \varepsilon_t, \qquad \varepsilon_t \sim \mathcal{N}(0, \sigma^2_{\varepsilon})
\ee
Here $y_t = X_t - \mu_t$ is the excess log-auction volume over the average; 
$X_t = \log(V^a_t)$ - log-auction volume; 
$\mu_t$ - average log-auction volume over the previous 20 days;
and $d_t$ - dummy variable for the quarterly option expiration days.
It allows us to calculate the option expiration dates multiplier $\eta_{a}=\exp(\beta d_t)$ for today's auction volume:
\be
V_t^{a}=\exp(\mu_t^a) \times \eta_{a}
\ee

For regular days, the behavior of the close auction volume is erratic and almost uncorrelated with trading activity until 30 minutes before close. For the special days specified above, there is a dependence between total daily volume of the continuous session and the close auction volume. In Figure 8 we plot the regression of the ratio of  the close auction volume to its geometric average  on ratio of the daily volume to its geometric average for the S$\&$P 500 representative sample for special days: 

\begin{figure}[!h]
    \caption{Close Auction Volume vs. Daily Volume for S$\&P$ 500 Subsample for Special Days}
  \hspace{2cm}
  \includegraphics[scale=0.9]{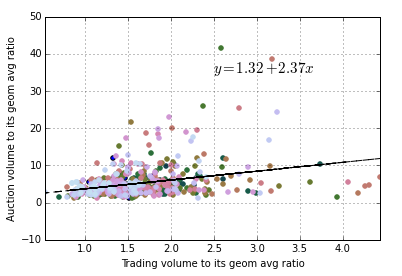}
\end{figure}

Better accuracy for the auction volume can be achieved using real-time auction imbalance information. This information becomes available only a few minutes before the close and can be combined with geometric average prior using the Bayesian technique described above. 

\section{Forming the Final Prediction}

The remaining daily volume $V_{rem}^D$ is the difference of estimated daily volume $V^D$ and volume traded so far $V(t)$: $V_{rem}^D(t)=V^D-V(t)$. It  is of main interest and is given by:
\begin{equation}
V_{rem}^D(t)= e^{\mu(t)} [1-\hat c (t)]
\end{equation}  
The estimated volume that is expected to be traded between time $t_1$ and $t_2$ is:
\begin{equation}
V({t_1,t_2})=(V_{rem}^D(t)+V(t)) [\hat c (t_2)-\hat c (t_1)] 
\end{equation}  

A human trader needs to know some practical numbers to set parameters of an execution algorithm, such as the expected urgency or the end time for her order.

Knowing the expected volume between $t_1$ and  $t_2$  $V({t_1,t_2})$ can help a trader to select the expected participation rate in Participation of Volume algorithm:
\begin{equation}
<\rho>_{t_1,t_2}=\frac{S}{V({t_1,t_2})}
\end{equation}  
here $S$ is the order size.  Alternatively, the predictor allows us to estimate the end time $t_2$ of the execution given a participation rate $\rho$:
\begin{equation}
t_2=\inf \{t: S=\rho\, V({t_1,t})\}
\end{equation}  

Similar analytics can be provided to VWAP traders who are interested in  u-curve analytics  (how strictly the VWAP schedule should be followed) and urgency recommendations. The dispersion information about the u-curve can be shown as well to offer reliability when using VWAP for a given stock.

\bigskip

\section{Conclusion}
We present a set of models relevant for predicting intra-day trading volume for equities. 
Instead of building a single complex model, we presented an approach where multiple simple models for historical daily volume and intra-day u-curve are calibrated separately and merged together by Bayesian formulae that automatically take into account the uncertainty of the model  input components. The models are calibrated using asymmetric error metrics (ALE) that give greater weight to overestimation error.

\bigskip

\section*{Acknowledgments}

%The authors gratefully thank Andrei Iogansen and Arun Verma for helpful discussions and suggestions.
We thank Andrei Iogansen for guarding the scientific rigor of our research and Arun Verma for his insightful feedback throughout this project.

Our gratitude goes to Arthur Umanski and Russell Langton for agile implementation and continuous support.

\section*{Mathematical Appendix}

Our main assumption is that the daily volume and intra-day bins follow a log-normal distribution. A random positive variable ${\displaystyle V}$  is log-normally distributed if the logarithm of ${\displaystyle V}$  is normally distributed: $X=log(V) \sim N(\mu,\sigma)$. For a normal variable $X$ with parameter $\theta = (\mu,\sigma)$ the likelihood of a sample of size $n$ is given by:
\begin{equation}
p(x|\theta)=\prod_{i=1}^n f_{\theta }(x_i)=(2 \pi)^{-n/2} \sigma^{-n} \exp\left(- \frac{1}{2\sigma^2} \sum_{i=1}^n (x_i-\mu)^2 \right)
\end{equation} 

Let's denote $\Sigma$ and $\bar x$ - empirical variance and mean of the sample.  Then taking into account the identity:
\begin{equation}
 \sum_{i=1}^n (x_i-\mu)^2=n\Sigma^2 + n (\bar x -\mu)^2
\end{equation} 

The log-likelihood is given by:
\begin{equation}
l_n(\mu,\sigma)=-\frac{n}{2} \log (2 \pi)-\frac{n}{2} \log(\sigma^2) -\frac{n \Sigma^2}{2 \sigma^2} -\frac{n (\bar x -\mu)^2}{2\sigma^2}
\end{equation} 

There are two flavors of Bayesian inference in our case: the distribution has known variance $\sigma^2$ but unknown mean $\mu$ and the distribution has unknown variance $\sigma^2$ and unknown mean $\mu$ (Murphy (2007)). 
% (Murphy (‎2007)).  

\subsection*{Bayesian Inference with Unknown Mean and  Known Variance}

Suppose the distribution has known variance $\sigma^2$ but unknown mean $\mu$.
Then likelihood as a function of $\mu$ is given by:
\begin{equation}
L_n(\mu)=p(x|\mu) \sim \exp \left( -\frac{n(\bar x-\mu)^2}{2\sigma^2} \right)
\end{equation} 
The conjugate prior is a Gaussian distribution with mean $\mu_0$ and variance $\sigma_0$. The probability density function of unnormalized prior $p_{prior}$
\begin{equation}
p(\mu) \sim \exp \left( -\frac{(\mu-\mu_0)^2}{2\sigma_0^2} \right) 
\end{equation} 
Equating  the product of $ L_n(\mu) p(\mu)$ to a probability density function (pdf) of a normal distribution  $N(\mu_p,\sigma_p)$ gives the mean $\mu_p$ and variance $\sigma_p^2$ of the posterior destribution $P_{posterior}=N(\mu_p,\sigma_p)$. 
The posterior mean of $\mu_p$ is expressed as a weighted average of the sample mean and the prior mean where the weights are proportional to precisions:

\begin{equation}
\mu_p=\frac{\frac{n \bar x}{\sigma^2} +\frac{\mu_0}{\sigma_0^2}}{\frac{n}{\sigma^2} +\frac{1}{\sigma_0^2} } \,\, 
\end{equation}  
and variance $\sigma^2$
\begin{equation}
\frac{1}{\sigma_p^2}=\frac{n}{\sigma^2} +\frac{1}{\sigma_0^2}  \,\, 
\end{equation}
here, $\bar x=\frac{1}{n} \sum_{i=1}^n x_i$.

Each observation increases the precision of the posterior distribution by the
precision $\lambda=\frac{1}{\sigma^2}$ of one observation. If we are interested only in inferences about the mean and 
the sample size is not too small, we can get a reasonable approximation to the posterior distribution by treating the standard deviation $\sigma$ as known and equal to the sample standard deviation $\Sigma_n$:
\be
\sigma^2 \approx \Sigma_n^2=\frac{1}{n}\sum_{i=1}^n (x_i-\bar x)^2
\ee
A more accurate representation of our knowledge should account for the unknown variance (or precision).

\subsection* {Bayesian Inference with Unknown Mean and Unknown Variance}
The conjugate prior for mean $\mu$ and precision $\lambda=1/\sigma^2$ is a normal gamma distribution:
\begin{multline}
p_{prior}(\mu,\lambda)=N(\mu|\mu_0,(\kappa_0 \lambda_0)^{-1}) Gamma(\lambda|\alpha_0,\beta_0)= \\
=\frac{1}{Z_{NG}} \lambda^{\alpha_0-1} \exp(-\beta_0 \lambda) \times \lambda^{1/2} \exp{\left(- \frac{1}{2} \kappa_0 \lambda (\mu-\mu_0)^2\right)}
\end{multline}

\begin{equation}
Z_{NG}=\frac{\Gamma(\alpha_0)}{\beta_0^{\alpha_0} }\left(\frac{2 \pi}{\kappa_0}\right)^{\frac 12}
\end{equation} 
here $\alpha_0, \beta_0,\mu_0,\kappa_0$ are hyper-parameters. 
The likelihood of n-observations with mean $x_n$ and variance $\Sigma_n$ 
\begin{equation}
L(x|\mu,\lambda)=\frac{1}{(2\pi)^{\frac{n}{2}}} \lambda^{n/2} \exp{\left( -\frac{\lambda}{2}\left[n(\mu-\bar x)^2+\sum_{i=1}^n (x_i-\bar x)^2 \right] \right) }
\end{equation} 
The posterior is given by normal gamma distribution with parameters  
$$
p(\mu, \lambda) = NG(\mu\lambda| \mu_n, \kappa_n, \alpha_n, \beta_n)
$$

\begin{equation}
\alpha_n=\alpha_0+\frac{n}{2},\,\,\, \beta_n=\beta_0+\frac{1}{2} \sum_{i=1}^n (x_i-\bar x)^2+\frac{\kappa_0 n (\bar x_n-\gamma_0)^2}{2(\kappa_0+n)}
\end{equation} 
\begin{equation}
\mu_n=\frac{\mu_0 \kappa_0+n \bar x}{\kappa_0+n}, \,\,\, \kappa_n=\kappa_0+n
\end{equation} 
The marginal distribution of mean $P(\mu|D)$ is given by t-distribution 
$$
P(\mu|D)=T_{2 \alpha_n}\left( \mu|\mu_n,\frac{\beta_n}{\alpha_n\kappa_n} \right) 
$$
\begin{equation}
\propto \left[ 1+\frac{(k_0+n)(\mu-\mu_n)^2}{\alpha_n \beta_n}\right]^{-\frac{n+\alpha_0+1}{2}}
\end{equation}

The estimation of posterior mean $\mu_p$ is a simple average between prior $\mu_0$  and average of $n$ observations $\bar x=\frac{1}{n}\sum_{i=1}^n x_i$.
\begin{equation}
\mu_p=\frac{\mu_0 \kappa_0+n \bar x}{\kappa_0+n},
\end{equation} 
here, $\kappa_0$  parameter is the effective size of the prior sample.

%%%%%%%%%%%%%%%%%%%%%%%%%%%%%%%%%
%%%%%%%%%%%%%%%%%%%%%%%%%%%%%%%%%
% REFERENCES
%%%%%%%%%%%%%%%%%%%%%%%%%%%%%%%%%
%%%%%%%%%%%%%%%%%%%%%%%%%%%%%%%%%


\newpage
\begin{thebibliography}{10}

%\bibitem{latexGuide} Almgren, R. and N. Chriss.
%"Optimal Execution of Portfolio Transactions."
%\emph{Journal of Risk}, Vol. 3 (December 2000),
%pp. 5-39

%\bibitem{latexGuide} Berkowitz, S.A., D.E. Logue, E.A.J.Noser.
%"The Total Cost of Transactions on the NYSE."
%\emph{The Journal of Finance}, Vol. 43, Issue 1 (March 1988),
%pp. 97-112.


%\bibitem{latexGuide} Bertsimas, D. and A.W. Lo
%"Optimal Contral of Execution Costs."
%\emph{Journal of Financial Markets}, Vol. 1 (1998),
%pp. 1-50.


%\bibitem{latexGuide} Bialkowski, J., S. Darolles, G. Le Fol. 
%"Improving VWAP Strategies: A Dynamical Volume Apprach."
%\emph{Journal of Banking and Finance}, Vol. 32, No. 9 (September 2008),
%pp. 1709-1722.

%\bibitem{latexGuide} Bialkowski, J., D. Mitchell, S.Tompaidis. 
%"Optimal VWAP Tracking."
%\url{http://www.nzfc.ac.nz/archives/2014/papers/programme/II-1d.pdf}

\bibitem{latexGuide} Brownlees, C.T., F. Cipollini, G.M. Gallo.
"Intra-Day Volume Modeling and Prediction for Algorithmic Trading."
\emph{Journal of Financial Econometrics}, 9 (2011),
pp. 489-518.

\bibitem{latexGuide} Calvori, F., F. Cipollini, G.M. Gallo.
"Go with the Flow: A GAS model for Predicting Intra-daily Volume Shares."
\url{http://local.disia.unifi.it/wp_disia/2014/wp_disia_2014_01.pdf}

\bibitem{latexGuide} Chiang, C.-H.
"Trading Volume and Option Expiration Date." (2009)
\url{https://editorialexpress.com/cgi-bin/conference/download.cgi?db_name=FEMES09&paper_id=634}

\bibitem{latexGuide} Chen, S., R. Chen, G. Ardell, B. Lin.
"End-of-Day Stock Trading Volume Prediction with a Two-Component Hierarchical Model."
\emph{The Journal of Trading}, (Summer 2011),
pp. 61-68.

%\bibitem{latexGuide}Chen, R., and T.B. Fomby.
%“Forecasting with Stable Seasonal Pattern Models with an Application to Hawaiian Tourism Data."
%\emph{Journal of Business and Economic Statistics}, Vol. 17, No. 4 (October 1999), 
%pp. 497-504.

\bibitem{latexGuide} Corredor, P., P. Lechon, R. Santamaria.
"Option-Expiration Effects in Small Markets: The Spanish Stock Exchange."
\emph{Journal of Futures Markets}, Vol. 21, Issue 10 (October 2001), 
pp. 905-928.

%\bibitem{latexGuide} Darolles, S., G. Le Fol. 
%"Trading Volume and Arbitrage."
%\emph{GSTF Business Review}, Vol. 3, Issue 3 (June 2014), 
%p. 30

\bibitem{latexGuide} Engle, R.F.
"New Frontiers for ARCH Models."
\emph{Journal of Applied Econometrics}, 17 (2002),
pp. 425-446.

\bibitem{latexGuide} Glukhov, V. S.
"Optimal trading in the presence of non-displayed liquidity"
%The Journal of Trading, Fall 2011, Vol. 6, No. 4, pp. 45-52 
\emph{The Journal of Trading}, Vol. 2007, No. 2 (Fall 2007),
pp. 30-37

\bibitem{latexGuide} Grubbs, Frank E.
"Sample criteria for testing outlying observations."
\emph{Annals of Mathematical Statistics}, Vol. 21, No. 1, (1950), pp. 27-58.

\bibitem{latexGuide} Gupta, A., S. Metia, P. Trivedi
"The Effects of Option Expiration on NSE volumne and prices." (2003)
\url{https://papers.ssrn.com/sol3/papers.cfm?abstract_id=619701}

\bibitem{latexGuide} Markov, V., S. Mazur, D. Saltz.
"Design and Implementation of Schedule-Based Trading Strategies Based on Uncertainty Bands"
\emph{The Journal of Trading}, Vol. 6, No. 4 (Fall 2011),
pp. 45-52 

\bibitem{latexGuide} Murphy, K., 
"Conjugate Bayesian analysis of the Gaussian distribution"
\url{https://www.cs.ubc.ca/~murphyk/Papers/bayesGauss.pdf}

\bibitem{latexGuide} Rashkovich, V.,  A. Verma.
"Trade Cost: Handicapping on PAR"
\emph{Journal of Trading},  Vol. 7, No. 4 (Fall 2012) 

\bibitem{latexGuide} Satish, V., A. Saxena, M. Palmer.
"Predicting Intraday Trading Volume and Volume Percentages."
\emph{The Journal of Trading}, Vol. 9, No. 3, (Summer 2014),
pp. 15-25.


\bibitem{latexGuide} Stone, G., T. Kingsley, G. Kan.
"The 12\% Rule For Asia’s Closing Auctions."
\url{http://fixglobal.com/home/the-12-rule-for-asias-closing-auctions/}


\bibitem{latexGuide} Swidler, S., L. Schwartz, R. Kristiansen.
"Option Expiration Day Effects in Small Markets: Evidence from the Oslo Stock Exchange."
\emph{The Journal of Financial Engineering}, Vol. 3, No. 2, (June 1994)

\bibitem{latexGuide} Vipul.
"Futures and options expiration-day effects: The Indian evidence."
\emph{ Journal of Futures Markets}, Vol. 25, Issue 11 (November 2005), 
pp. 1045-1065




\end{thebibliography}
\end{document}